\documentclass[aps,pra,onecolumn,groupedaddress,showpacs]{revtex4}

\usepackage{hyperref}
\usepackage{graphics}
\usepackage{amssymb,amsmath}
\draft
\usepackage{hyperref}
\usepackage{graphicx}
\newcommand{\Om}{\Omega}

\newcommand{\om}{\omega}
\newcommand{\al}{\alpha}

\newcommand{\ve}{\varepsilon}
\newcommand{\pa}{\partial}

\begin{document}

\title{Vector soliton of the Love wave in $ZnO/LiNbO_{3}$ structure}
\author{G. T. Adamashvili}
\affiliation{Technical University of Georgia, Kostava str.77, Tbilisi, 0179, Georgia.\\ email: $guram_{-}adamashvili@ymail.com.$ }

\begin{abstract}
A theory of an acoustic vector soliton  of the Love wave is constructed.
The nonlinear Love wave propagating along the interface between of a plane surface layer and the elastic semi-space under the condition of acoustic  self-induced transparency is investigated. A thin resonance transition layer containing paramagnetic impurity atoms or semiconductor quantum dots sandwiched between these two connected media. Explicit analytical expressions for the profile and parameters of the Love vector soliton are obtained.  It is shown that the properties of the nonlinear Love wave depends on the  parameters of the transition resonance layer, the connected elastic media and the transverse structure of the Love mode.
Numerical investigations of the Love vector soliton is executed for the parameters of the layered  structure $ZnO/LiNbO_{3}$ and surface acoustic waves which can be reached in current experiments.
\end{abstract}



\maketitle

\section{Introduction}

The propagation of acoustic nonlinear surface waves of the stable profile (solitons, breathers and vector solitons) is one of the most interesting demonstration of the acoustic nonlinearity in layered structures and in a variety of nanostructures. Depending of the character of the acoustic nonlinearity, the non-resonance and resonance mechanism of the formation of acoustic nonlinear waves is considered. In the case of non-resonance nonlinearity, which is expressed by means of anharmonically interacting phonons  its competition with the dispersion leads to the formation of non-resonance acoustic nonlinear waves \cite{Maradudin::1989,Adamashvili:Semiconductors:06,Sakuma::1984,Stevenson::03}.

Resonance acoustic nonlinear waves can be arise with the help of the resonance McCall-Hahn mechanism of the formation of nonlinear waves, i.e., when a nonlinear coherent interaction of an acoustic pulse with small concentration of paramagnetic impurities or quantum dots takes place and the conditions of acoustic  self-induced transparency (ASIT)
$\;\om T >> 1 $ and $ T<<T_{1,2}$ are fulfilled. Here, $\;\om$  and $T$ are the acoustic pulse frequency and width, respectively, while $T_{1}$ and $T_{2}$ are the longitudinal and transverse relaxation times of the resonance  impurity atoms or quantum dots. When the area of the acoustic pulse $\theta>\pi$, a soliton ($2\pi$ pulse) is formed, and for $\theta<<\pi$, small-area acoustic pulses ($0\pi$ pulses), for instance, breather can be propagated \cite{Shiren:PhysRev:70, Adamashvili:PhysLettA:12}.

The acoustic soliton (or breather) is  a single  acoustic pulse  propagates in such a way that it maintains its state. When these conditions are not fulfilled, we must to consider interaction between of two acoustic wave components at different frequencies as a bound state. Under this condition an acoustic vector soliton can be formed \cite{Adamashvili:PhysRevE:12}.

Nonlinear resonance  acoustic waves have been studied theoretically, as well as experimentally in various materials. Experimental considerations of acoustic resonance nonlinear waves have been carried out on paramagnetic crystals and in doped nanostructures, for instance $CaF_{2}:U^{4+},\; MgO:Fe^{2+},\; MgO:Ni^{2+},\;$ $KCl:OH^{-},\;$ $LiNbO_{3}:Fe^{2+}$ and $Mn$-doped $ZnS$ nanocrystals (see, for instance, Ref.\cite{Adamashvili:PhysLettA:12,Adamashvili:PhysRevE:12} and references therein).

The plane  acoustic nonlinear wave can be excited and propagates into the material perpendicular to the surface into the bulk \cite{Adamashvili:PhysLettA:12,Adamashvili:PhysRevE:12}, but on the surface or interface of the materials  the surface acoustic nonlinear wave can be generated  \cite{Adamashvili:Semiconductors:06,Kolomenskii::03,Hess:Ultrasonics:10}.
The most commonly used surface acoustic waves are Rayleigh and Love waves. The nonlinear surface acoustic waves have attracted much interest in the nano-acoustic systems and applications \cite{Kolomenskii::03,Capel::15,Devos::15,Kozhushko::08,Fu::2010}.

The shear horizontal (SH) polarized surface acoustic Love wave propagate in a layered structure consisting of a substrate and a layer on top of it \cite{Luthi::2004}. The following layered structures $ZnO/Quartz,\;ZnO/LiTaO_{3}, \; SiO_{2}/Quartz,\;SiO_{2}/LiTaO_{3}$ \cite{Ren-Chuan Chang::06}, $ZnO/LiNbO_{3}$ \cite{Shiou-Jen Jian::04}  are most often used in experimental studies of the Love-modes. The basic peculiarities of the surface Love waves are strong enhancement and spatial confinement of the energy of the elastic deformation  of the waves  inside of a thin guiding layer, while they decay evanescently in the substrate. The condition for the generate of Love modes is that the acoustic shear velocity in the layer is smaller than that of the substrate.

The surface acoustic solitons and breathers have been studied in many works \cite{Maradudin::1989, Adamashvili:Semiconductors:06,Kolomenskii::03,Hess:Ultrasonics:10,Adamashvili:PhysLettA:88,Sakuma::1984,Adamashvili:Sol.State. comm.:83,Capel::15,Devos::15,Kozhushko::08}. Recently, resonance vector soliton of the Rayleigh wave  have been investigated in Ref.\cite{Adamashvili:PhysRevE:16}, but resonance vector soliton of the Love wave have not been considered up to now.

The main goal of the present work is to theoretically investigate the formation of acoustic vector soliton of ASIT of the Love wave and the determination of the parameters and profile of the Love vector pulse. Numerical illustration is considered for a Love nonlinear wave with parameters of the layered  structure $ZnO/LiNbO_{3}$ which can be reached in current experiments.

\vskip+0.5cm

\section{Basic equations}

We study the formation of vector soliton of ASIT for a surface Love mode propagating along the interface  between of a plane surface layer acting as a guide (index \emph{l})  and the elastic substrate (index \emph{s}).  We assume that semi-space corresponding region $z\leqslant0$, and layer region $0< z\leqslant h$. A thin resonance transition layer  of thickness $d <<  \lambda$ containing a small concentration $n_{0}$ paramagnetic impurity atoms or semiconductor quantum dots with electron spin  $S=\frac{1}{2}$ sandwiched between these two connected media (at $z=0$), where $\lambda$ is the length of the Love wave. We assume that an external constant magnetic field $H_{0}$ is applied along the $z$-axis.
We shall consider a surface acoustic Love wave with width $T<<T_{1,2}$, frequency $\omega >>T^{-1}$, and wave vector $\vec{k}$, propagating along the positive $y$-axis. The Love wave is capable of cause excitations of the electron $S$ spins of the paramagnetic impurities or quantum dots \cite{Adamashvili:PhysLettA:87, Urbaszek:Rev.Mod.Phys.:2013}.

To take into account the magnetization caused by the presence of the resonance transition layer, the boundary conditions for Love modes at $z=h$,
\begin{equation}\label{Boun}
 \sigma^{(l)}_{xz}=0,
\end{equation}
and at $z=0$,
\begin{equation}\label{Boun2}
 \sigma^{(l)}_{xz}=\sigma^{(s)}_{xz}+ \sigma'_{xz},\;\;\;\;\;\;\;\;\;\;u^{(l)}_{x}=u^{(s)}_{x},
\end{equation}
where  $\sigma^{(l)}_{xz}$ and  $\sigma^{(s)}_{xz}$ are the stress tensor components in the layer and substrate, respectively. $\sigma'_{xz}$ is the contribution to the quantity $\sigma^{(s)}_{xz}$ caused by the presence of the transition layer with electron $S$ spins, $u^{(l)}_{x}$ and $u^{(s)}_{x}$ are the x-components of the deformation vector $\vec{u} (u_{x}, u_y, u_z)$ in connected media.

We will consider a Fourier-decomposition of the x-component of the deformation vector $u_{x}$ in connected media which is given by
\begin{equation}\label{Fu}
u_{x}(y,z,t)= \int  \tilde{u} (\Omega,Q) f(\Omega,Q,z) e^{i(Qy-\Om t)}
d\Om dQ.
\end{equation}
The Fourier amplitude which defines the transverse profile of the Love-mode has the following form:
\begin{equation}\label{kappa}
f(\Om,Q,z)=
\left \{%
\begin{array}{ccc}
        \tan [\kappa_{l}(\Om,Q)h] \sin[ \kappa_{l}(\Om,Q)z] +\cos [\kappa_{l}(\Om,Q)z],& \;\;\;\;\;0<z<h\\
    e^{\kappa_{s}z}, &\;\;\;\; z<0,
  \end{array}
\right \| %
\end{equation}
where
\begin{equation}\label{kapa}\nonumber\\
 \kappa_{l}^{2}= \frac{\Om^{2}}{c_{l}^{2}}-Q^{2},\;\;\;\;\;\;\;\kappa_{s}^{2}=Q^{2}-
 \frac{\Om^{2}}{c_{s}^{2}},
\end{equation}
$c_{s}$  and $c_{l}$ are the  transverse polarized sound velocities in the substrate and layer, respectively \cite{Landau:Theory of elasticity:80}.
The function $\tilde{u}(\Omega,Q)$ has to be determined.

The component of the deformation tensor at $z=0$ is given by
\begin{equation}\label{eps}
\varepsilon_{xz}(y,t)=\frac{1}{2}(\varepsilon^{+}+\varepsilon^{-})= \int   \tilde{\varepsilon}_{xz}(\Omega,Q)e^{i(Qy-\Om t)}d\Om dQ.
\end{equation}

We assume translational invariance in the $x$-direction, so that all field quantities  do not depend from the coordinate $x$, i.e. $\frac{\pa }{\pa x} \rightarrow 0.$

Substituting Eqs. \eqref{Fu} and \eqref{eps}   into  boundary conditions \eqref{Boun} and \eqref{Boun2}, we obtain the following nonlinear wave equation for  $\varepsilon_{xz}$ component of the deformation tensor at $z=0$:
\begin{equation}\label{epsw}
\int  \tilde{\varepsilon}_{xz}(\Omega,Q) F(\Omega,Q)      e^{i(Qy-\Om t)}d\Om dQ=-\frac{\sigma'_{xz}}{4\rho_{s} c^{2}_{s}},
\end{equation}
where
$$
F(\Omega,Q)=1-\frac{\rho_{l}c^{2}_{l}\kappa_{l}(\Omega,Q)}{\rho_{s}c^{2}_{s}\kappa_{s}(\Omega,Q)} \tan{[\kappa_{l}(\Omega,Q)h]},
$$
$$
\tilde{\varepsilon}_{xz}(\Omega,Q)=\frac{1}{2}\tilde{u} (\Omega,Q)\kappa_{s}(\Omega,Q),
$$
$\rho_{l}$  and $\rho_{s}$ are the  densities of the substrate and layer, respectively.

The Hamiltonian of the spin system of the paramagnetic impurities or quantum dots have the following form \cite{Adamashvili:PhysLettA:87, Weil :Electron Paramagnetic Resonance:2001, Landau:Quantum Mechanics:80}:
\begin{equation}\label{ham}
\hat{H}=\hat{H}_{Z}+\hat{H}_{sp},
\end{equation}
where  $\hat{H}_{Z}=\hbar \omega_{0} \hat{S}^{z}$  is the Hamiltonian of the Zeeman interaction,  $\hat{H}_{sp}=L \; \varepsilon_{xz} \hat{S}^{x} $ is the Hamiltonian of the interaction of the spin system with the phonons of the Love-mode, $L=\beta_{0} H_{0} F_{xzxz},\;$$\omega_{0}$ is the Zeeman frequency of the electron spin,
$\hbar$ is Planck's constant, $\beta_{0}$ is the Bohr magneton,  $F_{xzxz}$ is component of the spin-phonon coupling tensor, $\varepsilon_{xz}= \frac{1}{2}\frac{\partial u_{x}}{\partial  z}$. The function $\sigma'_{xz}=\frac{\partial <\hat{H}>}{\partial \varepsilon_{xz}}=L\;  S^{x}$ is defined from the Bloch equations, where quantities $S^{x,y,z}$ are the average values of the spin operators $\hat{S}^{x,y,z}$.

In the rotating-wave approximation the Hamiltonian $\hat{H}_{sp}$ can be transformed into the following form $$\hat{H}_{sp}=\frac{L}{4} (\varepsilon^{+}\hat{S}^{-}+\varepsilon^{-}\hat{S}^{+} ).$$
From the Hamiltonian \eqref{ham} we obtain the system of  the Bloch equations \cite{Weil :Electron Paramagnetic Resonance:2001,Landau:Quantum Mechanics:80}:
\begin{equation}\label{bloch}
\frac{\partial {S}^{+}}{\partial t}=i  \omega_{0} {S}^{+} -i\frac{L}{2 \hbar} {S}^{z}\varepsilon^{+},
$$$$
\frac{\partial {S}^{z}}{\partial t}=i\frac{L}{4 \hbar}({S}^{-}\varepsilon^{+}-{S}^{+}\varepsilon^{-}),
\end{equation}
where
$$
S^{\pm}=S^{x}\pm i S^{y}.
$$

The system of equations \eqref{bloch} we can transform to the slowly variables using the equations
\begin{equation}\label{spu}
\varepsilon_{xz}=\frac{1}{2} \sum_{l=\pm 1}\hat{E}_{l}Z_{l},\;\;\;\;\;\;\;\;\;\;\;\;\;
{S}^{\pm}=\pm i\rho^{\pm}Z_{\mp 1},
$$$$
\varepsilon^{\pm}=\hat{E}_{\mp 1}Z_{\mp 1},\;\;\;\;\;\;\;
Z_{l}= e^{il(ky -\om t)}.
\end{equation}

Substituting Eqs. \eqref{spu}  into \eqref{bloch} we obtain the Bloch equations for slowly envelope functions
$\hat{E}_{\pm 1}$ and $\rho^{\pm}$:
\begin{equation}\label{blochsw}
 \frac{\partial \rho^{+}}{\partial t}=i \Delta \rho^{+} - \frac{L}{2 \hbar}  {S}^{z}\hat{E}_{-1},
$$$$
\frac{\partial {S}^{z}}{\partial t}=\frac{L}{4 \hbar} (\rho^{-}\hat{E}_{-1}+\rho^{+} \hat{E}_{+1}),
\end{equation}
where $\Delta=\omega_{0}-\omega$.

The system of Eqs. \eqref{epsw}  and \eqref{blochsw} are the equations for ASIT for Love wave  and is the main object of our investigations. This system of equations can describe wide class of nonlinear resonance phenomena for surface Love mode.
For the solution of these equations, we will use the perturbation theory to study the evolution of the surface  acoustic vector soliton with two different frequency of oscillations for Love wave.

\vskip+0.5cm

\section{ Solution of the wave equation}

For the solution of the wave equation \eqref{epsw}, we present the function $F(\Om,Q)$ in the form of the series
\begin{equation}\label{Gser}
 F(\Om,Q)=F(\om,k)+(\Om-\om) \frac{\pa
F}{\pa \Om}|_{\Om=\om,Q=k}+(Q-k) \frac{\pa F}{\pa
Q}|_{\Om=\om,Q=k}+
$$$$+\frac{1}{2}[(\Om-\om)^{2} \frac{\pa^{2}
F}{\pa \Om^{2}}|_{\Om=\om,Q=k}
+2(\Om-\om)(Q-k)\frac{\pa^{2}
F}{\pa \Om \pa Q}|_{\Om=\om,Q=k}+(Q-k)^{2} \frac{\pa^{2}
F}{\pa Q^{2}}|_{\Om=\om,Q=k}]......
\end{equation}
where $\om$ and $k$ are the frequency and wave number of the carrier wave, respectively.

Substituting Eqs. \eqref{Gser} into the equation \eqref{epsw}, we obtain the equation
\begin{equation}\label{Gserx}
\int [F(\om,k)+(\Om-\om) F'_{\Om}+(Q-k) F'_{Q}+
(\frac{\Om^{2}}{2} -\om \Om +\frac{\om^{2}}{2})
F''_{\Om}+$$$$
+(\Om Q -\om Q- k \Om +k \om )F''_{\Omega,Q}+(\frac{Q^{2}}{2} -kQ +\frac{k^{2}}{2}) F''_{Q}]\tilde{\ve}_{xx}(\Om,Q)e^{i(Qy-\Om t)} d\Om
dQ=-\frac{\sigma'_{xz}}{4 c^{2}_{s} \rho_{s}},
\end{equation}
where
\begin{equation}\nonumber\\
F'_{\Om}=\frac{\pa F}{\pa
\Om}|_{\Om=\om,Q=k},\;\;\;\;\;\;\;\;\;\;\;F'_{Q}=\frac{\pa F}{\pa
Q}|_{\Om=\om,Q=k},
$$$$
F''_{\Om}=\frac{\pa^{2} F}{\pa
\Om^{2}}|_{\Om=\om,Q=k},\;\;\;\;\;\;\;\;\;\;\;F''_{Q}=\frac{\pa^{2} F}{\pa
Q^{2}}|_{\Om=\om,Q=k},\;\;\;\;\;\;\;\;\;\;\;F''_{\Omega,Q}=\frac{\pa^{2} F}{\pa
Q \pa \Om}|_{\Om=\om,Q=k}.
\end{equation}

Using the denotation \eqref{eps} the Eq.\eqref{Gserx} will be transformed to the following form
\begin{equation}\label{Gak}
(A+ iB \frac{\pa }{\pa t}-iC \frac{\pa }{\pa y}-\frac{F''_{\Om}}{2}  \frac{\pa^2 }{\pa t^2}
+F''_{\Omega,Q}\frac{\pa^2 }{\pa t \pa y}
-\frac{F''_{Q}}{2}  \frac{\pa^2 }{\pa y^2})\ve_{xz}
=-\frac{\sigma'_{xz}}{4 c^{2}_{s} \rho_{s}},
\end{equation}
where
\begin{equation}\nonumber\\
A=F(\om,k)-\om F'_{\Om} -k F'_{Q} +\frac{\om^{2}}{2}F''_{\Om} + k \om F''_{\Omega,Q} +\frac{k^{2}}{2} F''_{Q},
$$$$
B=F'_{\Om} -\om F''_{\Om} - k F''_{\Omega,Q},\;\;\;\;\;\;\;\;\;
C=F'_{Q} -\om F''_{\Omega,Q} -k F''_{Q}.
\end{equation}

Substituting Eqs. \eqref{spu} into the equation \eqref{Gak}, we obtain
\begin{equation}\label{Ga}
 \sum_{l=\pm 1} Z_{l} \{ a \frac{\pa }{\pa t}
  + ib \frac{\pa^{2} }{\pa t^2}
  -i d \frac{\pa^{2} }{{\pa t}{\pa y}}
  +F''_{\Omega,Q} \frac{\pa^{3} }{{\pa t^2}{\pa y}}
 - \frac{ F''_{\Om}}{2} \frac{\pa^{3}}{\pa t^{3}}
-\frac{F''_{Q}}{2}  \frac{\pa^{3} }{{\pa y^2} {\pa t}} \}{{\Theta}_{l}}
=-\frac{\sigma'_{xz}}{4 c^{2}_{s} \rho_{s}},
\end{equation}
where
\begin{equation}\nonumber\\
a=A  +   l(B \om + C k ) +\frac{1}{2} F''_{\Om} \om^{2} +  F''_{\Omega,Q} \om k + \frac{1}{2} F''_{Q} k^2,
 $$$$
b=B +l( F''_{\Om} \om +  F''_{\Omega,Q} k),\;\;\;\;\;\;\;\;\;\;\;
d =C  + l( F''_{\Omega,Q} \om +  F''_{Q} k),
$$$$
{\hat{E}_{l}}=\frac{\pa {{\Theta}_{l}}}{\pa t}.
\end{equation}

To further analyze of the Eq. \eqref{Ga} we make use of the multiple scale perturbative reduction method \cite{Taniuti::1973}, in the limit that $\Theta_{l}$ is of order $\epsilon $. This is the typical scaling for the coupled nonlinear Schrodinger equations and consequently,  would be the scaling for acoustic vector soliton. In this situation the function $\Theta_{l}(y,t)$ can be represented as in Refs. \cite{Adamashvili:Result:11, Adamashvili:Optics and spectroscopy:2012, Adamashvili:Physica B:12, Adamashvili:PhysRevE:12, Adamashvili:PhysLettA:2015}:
\begin{equation}\label{ttp}
\Theta_{l}= \sum_{\alpha=1} \varepsilon^\alpha {{\Theta}_{l}}^{(\alpha)}=\sum_{\alpha=1}^{\infty}\sum_{n=-\infty}^{+\infty}\varepsilon^\alpha
Y_{l,n} f_{l,n}^ {(\alpha)}(\zeta,\tau),
\end{equation}
where
$$
Y_{l,n}=e^{in(Q_{l,n}y-\Omega_{l,n}
t)},\;\;\;\zeta_{l,n}=\varepsilon Q_{l,n}(y-{v_g}_{l,n}
t),\;\;\;\tau=\varepsilon^2 t,\;\;\;
{v_g}_{l,n}=\frac{d\Omega_{l,n}}{dQ_{l,n}},
$$

At this it is assumed that the quantities $\Omega_{l,n}$, $Q_{l,n}$, and $f_{l,n}^{(\alpha)}$ satisfy the
inequalities for any $l$ and $n$:
\begin{equation}\label{rtyp}\nonumber\\
\omega\gg \Omega_{l,n},\;\;k\gg Q_{l,n},\;\;\;
\end{equation}
$$
\left|\frac{\partial
f_{l,n}^{(\alpha )}}{
\partial t}\right|\ll \Omega_{l,n} \left|f_{l,n}^{(\alpha )}\right|,\;\;\left|\frac{\partial
f_{l,n}^{(\alpha )}}{\partial y }\right|\ll Q_{l,n}\left|f_{l,n}^{(\alpha )}\right|.
$$
The quantities  $Q$, $\Omega$, $\zeta$ and $v_g$ depends from
$l$ and $n$, but for simplicity, we omit these indexes in equations where this will not bring about mess.

Substituting Eq.\eqref{ttp} into the Bloch equations \eqref{bloch}, we can determine the stress tensor component
\begin{equation}\label{sigma}
\sigma'_{xz}=i \frac{  L^{2}n_{0}}{8 \hbar} \int \frac{g(\Delta) d\Delta}{1+T^2 \Delta^2} \sum_{l=\pm 1} l Z_{l}(\ve^{1}{{\Theta}_{l}}^{(1)}+\ve^{2} {{\Theta}_{l}}^{(2)}+\ve^{3}{{\Theta}_{l}}^{(3)}- \ve^{3}  \frac{ L^2}{8\hbar^2 }\int \frac{\pa {{\Theta}_{l}}^{(1)}}{\pa t} {{\Theta}_{-l}}^{(1)}{{\Theta}_{l}}^{(1)}dt) +O(\ve^4),
\end{equation}
where $g(\Delta)$ is the inhomogeneous broadening function of the spectral line of the paramagnetic impurities or quantum dots.

After substitution of the expansion \eqref{ttp} into the equation  \eqref{Ga}, and taking into account  the explicit form of the envelope of the $\sigma'_{xz}$ component of the stress tensor \eqref{sigma},
we obtain dispersion law for the Love wave \cite{Landau:Theory of elasticity:80}:
\begin{equation}\label{dis}
\tan{[\kappa_{l}(\Omega,Q)h]}=\frac{\rho_{s}c^{2}_{s}\kappa_{s}(\Omega,Q)}{\rho_{l}c^{2}_{l}\kappa_{l}(\Omega,Q)}
\end{equation}
and nonlinear wave equation
\begin{equation}\label{eq 18}
\sum_{\alpha=1}^{\infty}\sum_{n=-\infty}^{+\infty}\varepsilon^\al  Z_{+1}
Y_{+1,n}   \{
{W}_{+1,n}
+\varepsilon J_{+1,n}
\frac{\partial }{\partial \zeta}
+\varepsilon^2 h_{+1,n}\frac{\partial }{\partial \tau}
+i\varepsilon^{2}H_{+1,n} \frac{\partial^{2} }{\partial \zeta^{2}}\}f_{+1,n}^{(\alpha)}
=$$$$i  \ve^{3} Z_{+1} \frac{\alpha_{0}  L^2}{8\hbar^2 }\int \frac{\pa {{\Theta}_{+1}}^{(1)}}{\pa t} {{\Theta}_{-1}}^{(1)}{{\Theta}_{+1}}^{(1)}dt) +O(\ve^4),
\end{equation}
where
\begin{equation}\label{eq 19}\nonumber\\
 W_{+1,n}=-in(n G'_{\Om}{\Omega}^{2} +G''_{\Omega,Q}  {\Omega}^{2}Q + \frac{G''_{\Om}}{2}   {\Omega}^{3}+\frac{G''_{Q}}{2}  Q^{2} \Omega+ G'_{Q}n Q \Omega - \frac{1}{n} \alpha_{0}),
$$$$
J_{+1,n}=-Q[ 2 G'_{\Om} n \Omega  v_g  + G'_{Q}n(Q v_g +\Omega)+  G''_{\Omega,Q}   \Omega ({\Omega} +  2   Q v_g )+ \frac{3 G''_{\Om}}{2} {\Omega}^{2}   v_g + \frac{G''_{Q}}{2}   Q(Q v_g +  2 \Omega)],
$$$$
h_{+1,n}=2G'_{\Om}n\Omega + G'_{Q}nQ  + 2 G''_{\Omega,Q} Q \Omega + \frac{3 G''_{\Om}}{2}   {\Omega}^{2}+ \frac{G''_{Q}}{2}   Q^{2},
$$$$
H_{+1,n}= Q^{2}[ G'_{\Om} v_g^{2}+  G'_{Q}  v_g + n G''_{\Omega,Q}     v_g ( 2  \Omega
+   Q v_g )+ n \frac{3 G''_{\Om}}{2}\Omega  {v_g}^{2}  +n \frac{G''_{Q}}{2}  ( 2Q v_g +\Omega)],
$$$$
\alpha_{0}=\frac{n_{0} L^{2}}{16 \hbar \rho_{s} c^{2}_{s}} \int \frac{g(\Delta)}{1+T^2 \Delta^2}d\Delta.
\end{equation}

To determine the values of $f_{l,n}^{(\alpha)}$, we equate to zero the
various terms corresponding to the same orders of $\varepsilon$. Following the standard procedure (see, for instance \cite{Taniuti::1973, Adamashvili:Result:11, Adamashvili:Optics and spectroscopy:2012, Adamashvili:Physica B:12, Adamashvili:PhysRevE:12, Adamashvili:PhysLettA:2015}),
we determine that, only the following components of $f_{l,n}^{(1)}$ can differ from zero: $f _{\pm 1,\pm 1}^{(1)}$ or $f_{\pm 1,\mp 1}^{(1)}$ and the following relation holds $J_{\pm 1,\pm 1}=J_{\pm 1,\mp 1}=0$.
The connections between the parameters $\Omega $ and $Q$ at $l=+1$ and $n^2=1$ has the following form
\begin{equation}\label{eq 20}
n (G'_{\Om}{\Omega}^{2}+ G'_{Q} Q \Omega -  \alpha_{0}) +G''_{\Omega,Q}  {\Omega}^{2}Q + \frac{G''_{\Om}}{2}  {\Omega}^{3}+\frac{G''_{Q}}{2}  Q^{2} \Omega=0
\end{equation}
and the equation for the quantity
\begin{equation}\label{eqq 21}
v_{g}=\frac{- n   G'_{Q}\Omega - G''_{Q} Q \Omega-G''_{\Omega,Q}  {\Omega}^{2}}{n  G'_{Q} Q  +2 n G'_{\Om}\Omega     +2 G''_{\Omega,Q}  {\Omega}Q
+ \frac{3}{2} G''_{\Om}   {\Omega}^{2} + \frac{G''_{Q}}{2}   Q^{2}  }.
\end{equation}

Substituting  Eqs.\eqref{eq 20} and \eqref{eqq 21} into Eq. \eqref{eq 18}, we obtain the coupled nonlinear Schrodinger equations for functions $\lambda_{\pm}= \varepsilon f_{+1,\pm 1}^ {(1)}$ that describe the connection between two components of the acoustic nonlinear Love pulse
\begin{equation}\label{nses}
i  (\frac{\partial \lambda_{\pm}}{\partial t}+ v_{\pm }\frac{\partial \lambda_{\pm}}{\partial
y})+ p_{\pm} \frac{\partial^{2} \lambda_{\pm}
}{\partial y^2}       + g_{\pm}|\lambda_{\pm}|^{2} \lambda_{\pm}
+ r_{\pm}|\lambda_{\mp}|^{2} \lambda_{\pm}=0,
\end{equation}
where
\begin{equation}\label{nses1}
v_{\pm}=v_{g;+1,\pm 1},\;\;\;\;\;\;\;\;\;\;
p_{\pm}= \frac{H_{+1,\pm 1}}{- h_{+1,\pm 1}Q^{2}},\;\;\;\;\;\;\;\;\;\;\;g_{\pm}=\frac{\alpha_{0}  L^2}{16\hbar^{2} h_{+1,\pm 1}},\;\;\;\;\;\;\;\;\;\;\;r_{\pm}=g_{\pm}(1-\frac{\Omega_{\mp}}{\Omega_{\pm}}),
$$
$$
\Omega_{+}=\Omega_{l=\pm1,n=\pm1},\;\;\;\;\;\;\;\;\;\Omega_{-}=\Omega_{l=\pm1,n=\mp1}.
\end{equation}

The coupled nonlinear Schrodinger equations \eqref{nses}  describes the  functions $\lambda_{+}$ oscillating with the frequency $\om+\Om_{+}$ and $\lambda_{-}$ describes the wave oscillating with frequency $\om-\Om_{-}$. The nonlinear connection between these two waves is governed by the terms $r_{+}|\lambda_{-}|^{2}\lambda_{+}$ and $r_{-}|\lambda_{+}|^{2}\lambda_{-}$. A stable profile solution of the equations \eqref{nses} is an acoustic vector soliton of the Love mode.

The simplest way to ensure the steady-state solution is to consider the envelope functions to depend on the time and space coordinate only through the coordinate $ \xi = t- \frac{y}{V_{0}}$, where $V_{0}$ is the constant vector pulse velocity. We will search the steady-state solutions of the Eqs.\eqref{nses} for the envelope function of the Love mode in the following form \cite{Adamashvili:Result:11, Adamashvili:Optics and spectroscopy:2012, Adamashvili:Physica B:12, Adamashvili:PhysRevE:12}:
\begin{equation}\label{eq12q}
\lambda_{\pm}(y,t)=A_{\pm}\; S( \xi )e^{i\phi_{\pm}},
\end{equation}
where $\phi_{\pm}=k_{\pm} y- \omega_{\pm} t$ are the phase functions, $A_{\pm},\;$$\;k_{\pm}$ and $\omega_{\pm}$ are all real constants. The functions $ e^{i\phi_{\pm}}$ are  slow in comparison with oscillations of the pulse and consequently, the inequalities
\begin{equation}\label{eq12a}\nonumber\\
k_{\pm}<<Q_{\pm },\;\;\;\;\omega_{\pm}<<{\Omega}_{\pm }
\end{equation}
are satisfied.

Substituting Eqs.\eqref{eq12q}  into Eqs.\eqref{nses}   after integration we obtain the steady-state solutions:
\begin{equation}\label{eq12t}
\lambda_{\pm}= \frac{A_{\pm}}{b T} sech (\frac{t- \frac{y}{V_{0}}}{T}) e^{i\phi_{\pm }},
\end{equation}
where
\begin{equation}\label{rt16}
b^{2}=V_{0}^{2} \frac{A_{+}^{2}g_{+}+A_{-}^{2}r_{+}}{2p_{+}},\;\;\;\;\;\;\;\;\;
T^{-2}=V_{0}^{2}\frac{v_{+}k_{+}+k_{+}^{2}p_{+}-\omega_{+}}{p_{+}}.
\end{equation}
Substituting Eq.\eqref{eq12t}  into Eqs. \eqref{spu} and \eqref{ttp}, we obtain for the $\varepsilon_{xz}$ component of the deformation tensor at $z=0$:
\begin{equation}\label{eq17}
\varepsilon_{xz}(y,t) = \frac{1}{b T} sech(\frac{t-\frac{y}{V_{0}}}{T})\{ (\Omega_{+}+\omega_{+}) A_{+} \sin[(k+Q_{+}+k_{+})y -(\om +\Omega_{+}+\omega_{+}) t]$$$$
- (\Omega_{-}-\omega_{-})A_{-} \sin[(k-Q_{-}+k_{-})y -(\om -\Omega_{-}+\omega_{-})t]\},
\end{equation}
where the relations between the parameters $A_{\pm},\;$$\omega_{\pm}$ and $k_{\pm}$ have the form
\begin{equation}\label{rt16a}
A_{+}^{2}=\frac{p_{+}g_{-}- p_{-}r_{+}}{p_{-}g_{+}-p_{+}r_{-}}A_{-}^{2},
$$
$$
\omega_{+}=\frac{p_{+}}{p_{-}}\omega_{-}+\frac{V^{2}_{0}(p_{-}^{2}-p_{+}^{2})+v_{-}^{2}p_{+}^{2}-v_{+}^{2}p_{-}^{2}
}{4p_{+}p_{-}^{2}},
$$
$$
k_{\pm}=\frac{V_{0}-v_{\pm}}{2p_{\pm}}.
\end{equation}

\begin{widetext}

\begin{figure}
\includegraphics[width=0.88\textwidth]{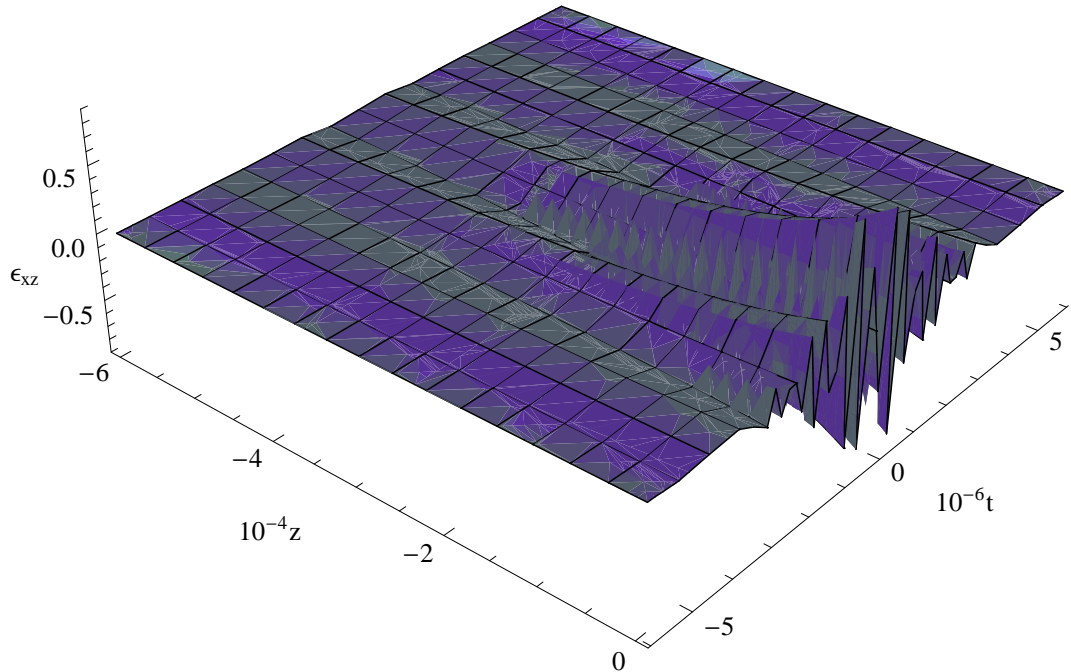}
\caption{(Color online)  Plot of the $\varepsilon_{xz}$ (in arbitrary units) component of the  deformation tensor  at a fixed value of the $y$ coordinate showing the two-dimensional vector soliton of the Love wave in $ZnO/LiNbO_{3}$ layered structure.  At $z=0$, the profile of the vector soliton corresponds to the solution of Eq. \eqref{eq17}.}
\label{fig1}
\end{figure}

\end{widetext}

The Eq.\eqref{eq17} is two-component vector soliton  solution  for the  $\varepsilon_{xz}$ component of the deformation tensor of the acoustic Love pulse. In expression \eqref{eq17}  the functions  $ \sin[(k+Q_{+}+k_{+})y -(\om +\Omega_{+}+\omega_{+}) t]$ and $\sin[(k-Q_{-}+k_{-})y -(\om -\Omega_{-}+\omega_{-})t]$ indicates of the two different frequencies of oscillations.

\vskip+0.5cm

\section{Conclusion}

In summary we have shown that in the propagation of the Love wave on the interface of the two elastic media with transition resonance monolayer containing an impurity atoms or quantum dots under the condition of ASIT  the vector soliton of the Love wave can arise. The explicit analytical expressions for the profile and parameters  of the acoustic two-component vector soliton of the Love wave are given by Eqs.\eqref{eq17}, \eqref{eqq 21}, \eqref{nses1}, \eqref{rt16} and \eqref{rt16a}.
The dispersion equation and the relations between quantities $\Omega_{\pm }$ and $Q_{\pm }$ are given by Eqs.\eqref{dis} and \eqref{eq 20}, respectively. The transverse profile of the Love-mode is given by Eq.\eqref{kappa}.

From these equations we can see that the properties of the nonlinear Love wave depends on the  parameters of the transition resonance layer, the connected elastic media and the transverse structure of the Love mode.

In the present work we have used the reduction perturbation expansion for  the Love wave under the condition of the ASIT to obtain the resonance Love-mode vector soliton  with two different  (sum $\om +\Omega_{+}$ and difference $\om -\Omega_{-}$ ) frequencies of oscillations.

Using typical parameters for the pulse,  $ZnO/LiNbO_{3}$ layered structure, and the paramagnetic impurities $Fe^{2+}$  \footnote {Parameters for the numerical simulation: $\omega =2\pi \times 10^{10} {\rm Hz}$, $T=2.35 \times 10^{-6}$ s, $n_0 = 10^{22} \, {\rm cm^{-3}},\;$ $h=6\times 10^{-7}\; cm,\;$ $c_{s}=4.478\times 10^{5}$ cm/s, $c_{l}=2.57 \times 10^{5}$ cm/s, $H_{0}=2050 \;F,\;$ $\rho_{s}=4.65\;g/cm^{3},$ $\rho_{l}=3.58\;g/cm^{3},\;$ $F_{xzxz}=725 \, {\rm cm^{-1}}$,
full-width half-maximum inhomogeneous broadening $T_{2}^* =\pi g(0)=3\times 10^{-9} \, s.$}, we can construct a plot of the $\varepsilon_{xz}$  component of the deformation tensor for a two-component acoustic vector soliton of the Love wave (shown in Fig.1 for a fixed value of the $y$ coordinate).

We have to note that zinc oxide ($ZnO$) is suitable material for the guiding layer of the Love-modes\cite{Fu::2010} and lithium niobate ($LiNbO_{3}$) is very convenient substrate material for surface acoustic waves, as well as often used for experimental investigations of the nonlinear acoustic waves \cite{Samartsev:JETP Lett.:1974, Ilinskii::2014}.

The results of this theoretical study of resonance two-component acoustic vector soliton of the Love wave, together with those obtained in Refs.\cite{Adamashvili:Sol.State. comm.:83, Adamashvili:PhysLettA:87} for one-component solitons and breathers  provide a more complete physical description of the propagation of resonance nonlinear surface acoustic Love waves of ASIT in layered structure.

The presented analytical and numerical results give grounds to hope that the two-component vector soliton of the Love mode can be observed experimentally.   Such investigation will be informative not only for the study of resonance vector solitons of the Love waves,  but also will be important for applications in acoustic devices based on the surface acoustic Love waves.

\end{document}